# Vapour sensing properties of graphene-covered gold nanoparticles

Gábor Piszter[a,b], Krisztián Kertész[a,b], György Molnár[a], András Pálinkás[a,b], András Deák[a], and Zoltán Osváth[a,b,*]

[a]*Institute of Technical Physics and Materials Science, MFA, Centre for Energy Research, Hungarian Academy of Sciences, 1525 Budapest, P.O. Box 49, Hungary*

[b]*Korea-Hungary Joint Laboratory for Nanosciences (KHJLN), 1525 Budapest, P.O. Box 49, Hungary*

**Abstract**

We investigated the vapour sensing properties of different graphene-gold hybrid nanostructures. We observed the shifts in the optical spectra near the local surface plasmon resonance of the gold nanoparticles by changing the concentration and nature of the analytes (ethanol, 2-propanol, and toluene). The smaller, dome-like gold nanoparticles proved to be more sensitive to these vapours compared to slightly larger, flat nanoparticles. We investigated how the optical response of the gold nanoparticles can be tuned with a corrugated graphene overlayer. We showed that the presence of graphene increased the sensitivity to ethanol and 2-propanol, while it decreased it towards toluene exposure (at concentrations ≥30%). The slope changes observed on the optical response curves were discussed in the framework of capillary condensation. These results can have potential impact on the development of new sensors based on graphene-gold hybrids.

*Corresponding author. E-mail address: zoltan.osvath@energia.mta.hu

**1 Introduction**

Graphene has fascinating mechanical, thermal, and electrical properties, which make this two-dimensional crystal with sp$^2$-hybridized carbon atoms arranged in a honeycomb lattice very attractive for several applications in the field of nanotechnology [1,2]. It is an ultrasensitive material for detecting gas molecules due to the large surface area [3] and the capability of all carbon atoms in graphene to interact with adsorbed molecules. The adsorption of targeted chemical species induces changes in the conductivity of the graphene sheet [4,5], which can be monitored by an appropriate sensing device. Several reviews on graphene-based gas/vapour sensors utilizing different operating principles were published recently [6,7,8,9].

Nobel metal nanoparticles (NPs) have also been in the focus of considerable interest for possible sensing applications [10]. Here, the sensing properties are determined either by the shift induced in the localized surface plasmon resonance (LSPR) of the NPs, or due to stronger light-matter interactions (surface enhanced Raman scattering - SERS). The LSPR is manifesting as absorption and scattering peaks as well as strong near-field enhancements, which occur when the incident light frequency matches the collective oscillation frequency of conduction band electrons. The wavelength of the LSPR peak depends sensitively on the shape, size, and neighbourhood conditions of the metal NPs [11,12]. Therefore, it can be used for detection of adsorbed volatile organic compounds or many other biochemical molecules by visible spectroscopy measurements of the LSPR peak shift [see Ref. [13] for a review].

The synthesis of graphene-metal nanoparticle hybrid materials has been one of the many efforts dedicated to enhance the sensing properties of graphene based sensors [14,15,16,17]. Besides combining the unique properties of graphene and the advantages of metal NPs, these hybrid nanostructures can display synergistic effects or novel functions as well [18]. The role of metal-graphene hybrid nanostructures in promoting the performance of LSPR sensors was discussed very recently in a focused review [19]. In particular, graphene/gold nanoparticle hybrids were mainly studied for electrochemical [20] or SERS-based sensing [21]. The chemical sensing properties of graphene covered optical nano-antenna arrays made of gold were also tested by exposure to vapour phase organic solvents and measuring the shift of the resonance peak [22]. However, the fabrication of such nano-antenna arrays on a larger scale must be time consuming and involves precise e-beam lithography technique. In this work we use a simple way for the preparation of large area graphene-covered gold nanoparticles. We apply a corrugated graphene overlayer obtained by annealing at moderate temperatures. The use of such rippled graphene can be beneficial in terms of chemical activity, as the crests and troughs of graphene ripples can form active sites for the adsorption of different molecules [23]. We show that the NPs display pronounced optical response upon exposure to organic vapours (ethanol, 2-propanol, toluene), and that the corrugated graphene overlayer can improve the selectivity.

**2 Experimental**

*2.1 Preparation of graphene/gold nanoparticle hybrid structures*

Gold layers of 5 and 10 nm thickness were evaporated onto 285 nm-SiO$_2$/Si substrate at a rate of 0.1 nm s$^{-1}$ and background pressure of 5 × 10$^{-7}$ mbar. The substrate was held at room temperature during evaporation. The deposited gold films were transformed into nanoparticles by annealing at 400 °C under Ar atmosphere for 30 minutes.

Graphene was grown by chemical vapour deposition on a copper foil, as described in our recent publication [24]. In order to transfer large-area graphene onto the gold NPs, we used thermal release tape, and an etchant mixture consisting of a CuCl$_2$ aqueous solution (20%) and HCl (37%) in 4:1 volume ratio. After etching the copper foil, the tape holding the graphene was rinsed in distilled water, then dried and pressed onto the nanoparticle-covered SiO$_2$ substrate. The tape/graphene/Au NPs/SiO$_2$/Si sample stack was placed on a hot plate and heated slightly above the release temperature of the tape (95 °C). Graphene-covered Au NPs were obtained by simply removing the tape. The graphene-covered samples were further annealed at 400 °C for 60 minutes to improve the adhesion of graphene to the NPs. Structural characterization was performed by a MultiMode 8 atomic force microscope (AFM) from Bruker, operating in tapping mode under ambient conditions.

*2.2 Optical spectrometry and the vapour sensing setup*

The optical reflectance and the vapour sensing measurements were conducted by fixing the samples in an air-proof aluminium box covered with a quartz glass window to provide UV transmittance. For the illumination of the samples an Avantes AvaLight DH-S-BAL light source was used and the initial reflectance of the samples in air was measured by collecting the specular reflected signal (measured under 15°) with an Avantes HS 1024*122TEC spectrometer (Avantes BV, Apeldoorn, The Netherlands). During the vapour sensing measurements, three types of volatile vapours were passed through the cell's gas inlet and exhausted through the outlet: ethanol, 2-propanol (IPA), and toluene (analytical grade, VWR International Ltd, Radnor, PA, USA). The vapour concentration was set by switching digital mass flow controllers (Aalborg DFC, Aalborg Instruments & Controls, Inc., Orangeburg, NY, USA) to let pass synthetic air (Messer, 80% $N_2$, 20% $O_2$) and saturated volatile vapours from gas bubblers in the required ratio. A constant gas flow of 1000 ml/min through the cell was maintained during the measurements. Vapour sensing experiments were carried out by changing the concentration and the type of test vapours while monitoring the spectral variations in time: 20 s mixture flow was followed by 60 s of synthetic air flow, to purge the cell. A 60 s purging was used to recover the initial reflectance value before the introduction of the next vapour mixture.

## 3 Results and discussion

The structure of the samples is shown in Fig. 1, as measured by AFM. The annealing of the 5 nm gold film resulted in dome-like nanoparticles with average height of 15 nm and lateral dimensions ranging from 10 to 60 nm (Fig. 1a). On the other hand, when annealing the 10 nm gold film, we obtained mostly elongated, rather flat nanoparticles with average height of 22 nm and lengths in the range of 50 – 300 nm (Fig. 1b). To simplify the notations, we will refer to these NPs as "5 nm" and "10 nm" Au NPs, respectively. The AFM image of the graphene-covered dome-like NPs is shown in Fig. 1c. Note that the annealing performed after graphene transfer (see the Experimental section) induces extended ripples on the 10 nm scale in the graphene supported by the "5 nm" NPs (see also the height profile in Fig. 1d). These ripples can be characterized by a $h/R$ ratio of 0.1÷0.2, where $h$ is the ripple height and $R$ is the equivalent radius of a nanotube-like elongated ripple. Such ripples can induce local strain values of order $(h/R)^2 \approx 1-4\%$.

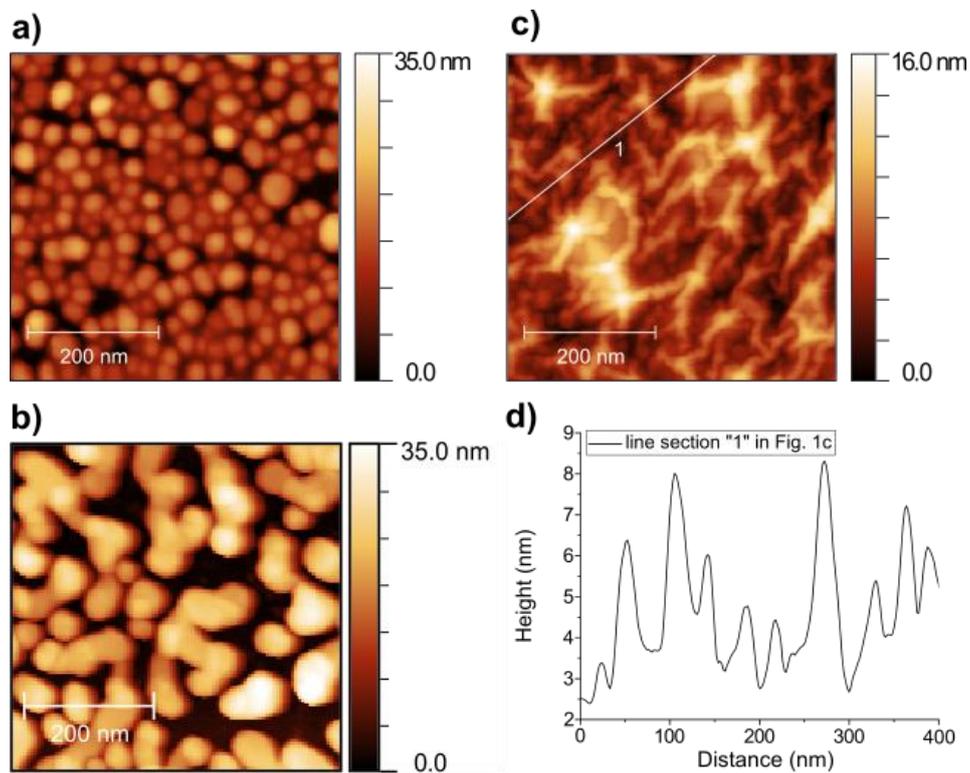

**Figure 1.** *Tapping mode AFM images of Au NPs obtained from annealing (a) 5 nm gold layer on $SiO_2$, (b) 10 nm gold layer on $SiO_2$. (c) Au NPs as in a), but covered with graphene. (d) Height profile along the line section "1" in c).*

The optical reflectance spectra of the samples are shown in Fig. 2b. Here, bare $SiO_2$ was used as a reference, meaning that all spectra were divided by the measured reflectance spectrum of the bare $SiO_2$ surface. The measurement was carried out by fixing a sample inside the vapour sensing cell and artificial air atmosphere was applied. Specular illumination and light collection were applied with 15° degrees between the two optical fibres (Fig. 2a). The reflectance spectrum of the "5 nm" NPs shows a prominent minimum around 625 nm (Fig. 2b, green line), which is attributed to the plasmon-coupling between the individual dome-like nanoparticles due to the small inter-particle separation (also see Fig. 1a) [25]. The plateau around 540 nm is the contribution from the native dipole mode of the dome-like particles [Fig. S1, ESI]. A small blue shift of 4.5 nm is observed in the reflectance minimum when graphene is transferred onto these NPs (Fig. 2b, red line). We showed recently [26], that annealing at moderate temperatures decreases the graphene-Au NP separation, increasing thus the interaction, and the blue shift is the result of p-type doping of graphene from the NPs [27]. We note that a blue shift in the LSPR occurs also if the inter-particle distances increase [28]. However, this scenario is not confirmed by AFM measurements [see Fig. S3, ESI].

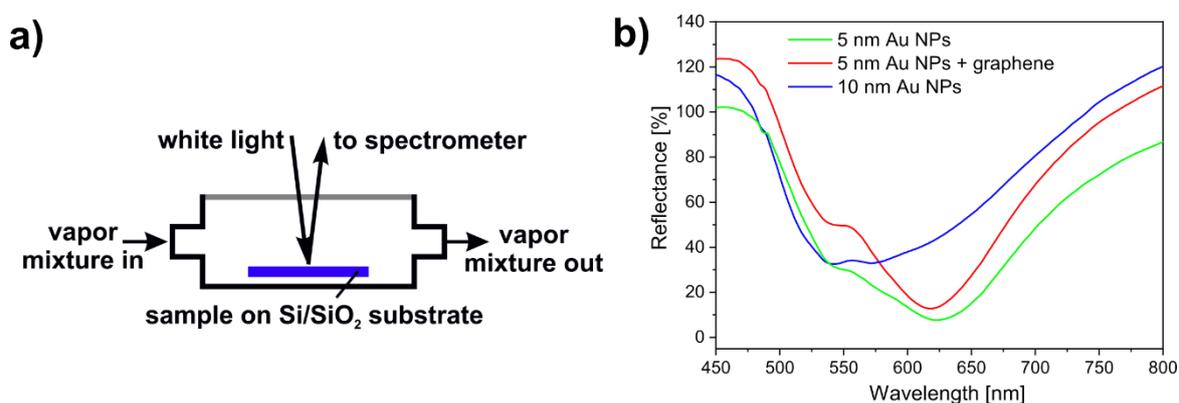

*Figure 2. (a) Schematic drawing of the aluminum cell used in the vapour sensing measurements. (b) Optical reflectance spectra of Au NPs and the annealed graphene/"5 nm" Au NPs samples measured in air.*

The larger, more irregularly shaped nanoparticles of the "10 nm" NPs sample (Fig. 2b, blue line) display a broad reflectance minimum due to the combination of shape anisometry (which itself causes broadening) and plasmon coupling [29,30]. Nevertheless, the characteristic dipole LSPR mode around 540 nm can be also observed, as expected. The spectra in Fig. 2b show the initial reflectance of the samples in air. These spectra were measured using blank $SiO_2$/Si wafer as a reference, and in turn, they were used as references in the vapour sensing experiments.

The spectral change during vapour exposure was characterized by dividing the actually measured spectrum with the respective reference.

A direct comparison between the bare Au NP samples is given in Figure 3, where we show the optical reflectance change of the two Au NP samples during vapour exposure. Three different vapours were used independently: ethanol, 2-propanol (IPA), and toluene. In each case, the organic vapour (33%) was diluted with artificial air (66%). Note that the same vapour concentration caused different change of the spectral amplitude for the two samples.

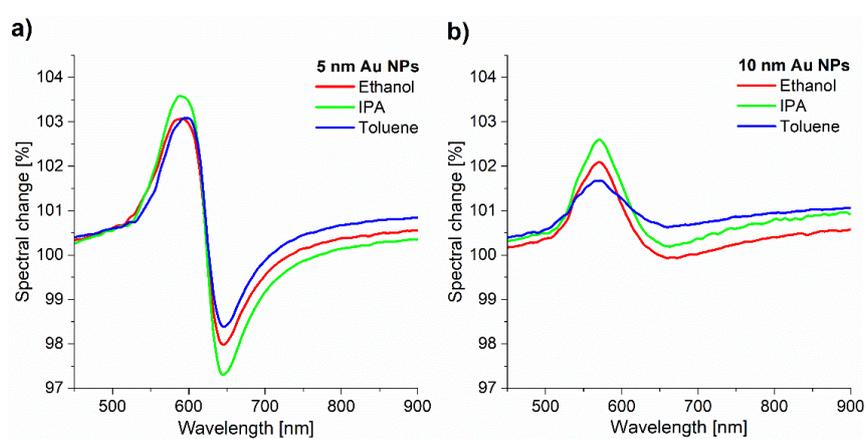

*Figure 3. Optical reflectance change of the (a) "5 nm" Au NPs, and (b) "10 nm" Au NPs during exposure to three different vapours (33%): ethanol, IPA, and toluene. The change is relative to the initial spectrum of each sample in artificial air.*

Generally, larger nanoparticles produce stronger near-field related optical effects (eg. SERS, or refractive index sensitivity). In the case of the "10 nm" Au NPs, however, the large anisometry in both size and shape resulted only in a much broader reflection minimum (Fig. 2b), and thus a less sensitive response to vapours (Fig. 3b). In contrast, the Au NPs of the "5 nm" sample gave a better optical response (Fig. 3a). This can be explained by the lower size dispersion, more uniform shape, and the strong coupling between the NPs due to the high nanoparticle density. The well-defined coupled mode around 625 nm (Fig. 2b) is therefore better suited to study LSPR shifts. Thus, in the following we used only the "5 nm" Au NPs sample for further sensing experiments. Concentration-dependent measurements were carried out using 10% concentration steps from artificial air to saturated vapours. Figure 4a shows the corresponding spectral changes of graphene covered "5 nm" Au NPs sample during ethanol exposure. As expected, the optical response increases with the vapour concentration, as higher number of adsorbed molecules increase more the effective refractive index of the medium (see Fig. S2, ESI). The maximal values of the response peaks observed at 580 nm are plotted as a function of ethanol vapour concentration in Fig. 4b. These values are compared to the maxima of the response curves measured upon ethanol exposure on bare "5 nm" Au NPs. Similar data are extracted from IPA and toluene exposure and plotted in Fig. 4c and Fig. 4d, respectively. In the case of bare Au NPs, the spectral change reaches 132% (relative to the initial reflectance in artificial air) for saturated IPA, 126% for saturated toluene, and 118% for saturated ethanol vapours. These values do not correlate directly with the refractive indices (*n*) of the corresponding solvents, since *n*(IPA) = 1.376, *n*(toluene) = 1.496, and *n*(ethanol) = 1.361. Note that, due to the larger refractive index, toluene should produce the largest spectral shift (Δ*λ*), according to the following relation [31,32]:

$$\Delta\lambda = m(n - n_0)\left[1 - e^{-2d/l_d}\right], \tag{1}$$

where $m$ is the refractive index response of the Au NPs, $n_0$ is the refractive index of air, $d$ is the thickness of the adsorbate, and $l_d$ is the characteristic field decay length. Nevertheless, taking into account the vapour pressures (at 20 °C) of the three solvents, 4.4 kPa, 3.0 kPa, and 5.9 kPa, respectively [33], we can say that at a given volumetric mixing ratio the toluene vapour contains smaller number of molecules, than either the IPA or ethanol vapours, which can explain the similar optical response for toluene and IPA: the smaller number of molecules compensates for the larger refractive index. Furthermore, graphene has a complex refractive index of about 2.65+1.27i [34,35]. We expect that its presence as a top layer on Au NPs should further increase the spectral change in all cases. However, this is not what we have observed.

The graphene-covered sample shows higher sensitivity to ethanol (Fig. 4b), and to IPA vapours (Fig. 4c), compared to the bare Au NPs, while it is less sensitive to toluene (Fig. 4d). The spectral change increases from 132% to 150% for saturated IPA, decreases from 126% to 118% for saturated toluene, and increases from 118% to 138% for saturated ethanol vapours.

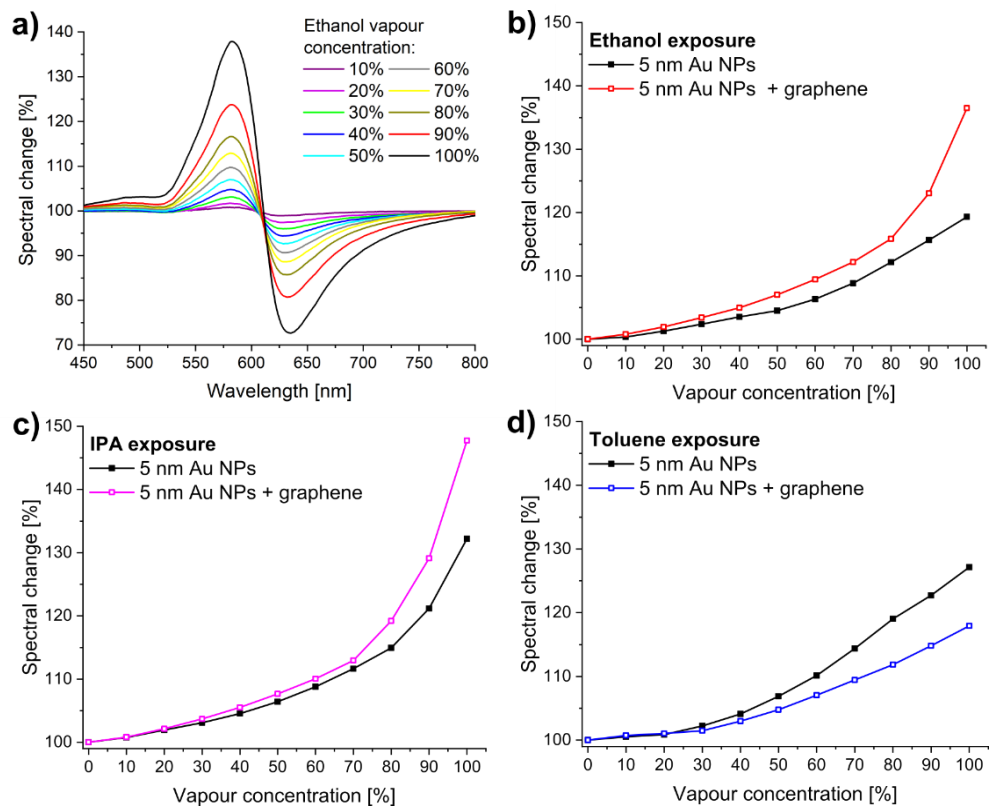

*Figure 4. Concentration-dependent vapour sensing measurements carried out using 10% concentration steps from artificial air (0%) to saturated vapours (100%). (a) The spectral change of graphene/"5 nm" Au NPs sample during ethanol exposure (the reference is the initial reflectance in artificial air). (b)-(d) The maximal spectral change was plot as a function of vapour concentration (see also Fig. 5). The graphene-covered sample shows higher sensitivity to (b) ethanol, and (c) IPA vapours, compared to the bare Au NPs, while it is less sensitive to (d) toluene.*

The time responses of bare Au NPs and graphene-covered Au NPs for vapour exposures are shown in Fig. 5a-c. In every case, the initial spectral response is a pronounced sharp, linear increase (decrease) as the vapour flow starts (stops), which infers response (and recovery) times as low as 2-3 seconds. These short response and recovery times are observed for graphene-covered Au NPs as well. There is no significant shift of the baseline at lower concentrations (Fig. 5d), while some slight memory effect can be observed at concentrations higher than 40%. The numerical values of the spectral changes shown in Fig. 5a-c are summarized in Table 1.

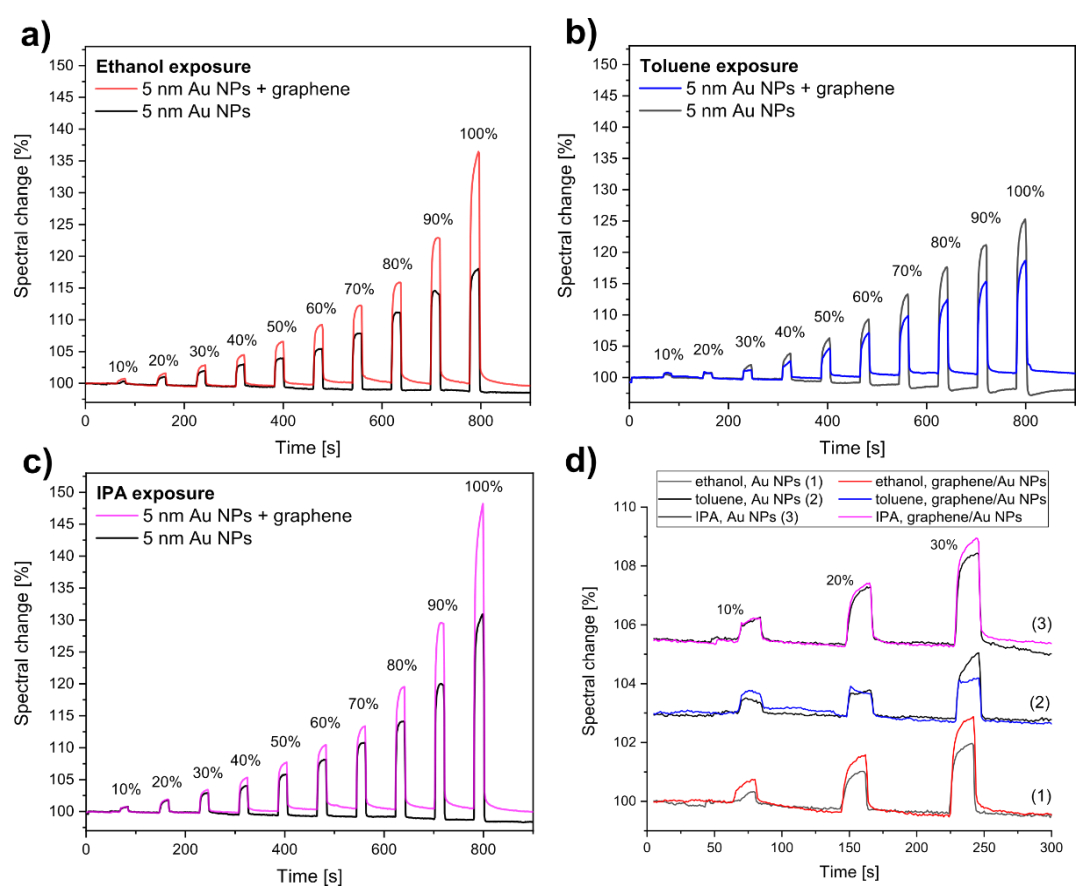

*Figure 5.* Vapour sensing responses of both bare "5 nm" Au NPs (black lines) and the graphene/"5 nm" Au NPs sample (coloured lines), averaged on 20 nm interval around the maximal spectral change. Exposures to (a) ethanol (b) toluene, and (c) IPA were done for 20 s, followed by purging in synthetic air for 60 s. Vapour concentration steps of 10% were used. Lower concentration (≤30%) spectral responses are magnified in (d). The spectral curves of toluene and IPA exposures were shifted vertically for clarity.

Considering the time responses at low vapour concentration (10%), we find that there are significant differences between the optical responses of graphene-covered and non-covered Au NPs. We measure two times larger spectral change for ethanol and 50% increase for toluene with the graphene-covered sample. In comparison, the response for IPA is similar for both graphene-covered and bare Au NPs. Interestingly, at this lower concentration the graphene-covered Au NPs give a more sensitive response for toluene, compared to bare NPs. This becomes less sensitive only at higher concentrations (≥30%). We think that this change in the sensitivity is related to capillary condensation, as described in the next paragraph.

| Vapour concentration (%) | Spectral change relative to baseline (%) | | | | | |
|---|---|---|---|---|---|---|
| | Au NPs, ethanol | gr/Au NPs, ethanol | Au NPs, toluene | gr/Au NPs, toluene | Au NPs, IPA | gr/Au NPs, IPA |
| 10 | 0.35 | 0.78 | 0.53 | 0.72 | 0.76 | 0.80 |
| 20 | 1.27 | 1.95 | 0.87 | 1.03 | 1.95 | 2.14 |
| 30 | 2.37 | 3.40 | 2.22 | 1.49 | 3.10 | 3.69 |
| 40 | 3.52 | 4.97 | 4.11 | 2.98 | 4.55 | 5.52 |
| 50 | 4.50 | 7.01 | 6.88 | 4.76 | 6.40 | 7.66 |
| 60 | 6.34 | 9.44 | 10.14 | 7.05 | 8.79 | 10.05 |
| 70 | 8.84 | 12.19 | 14.39 | 9.45 | 11.63 | 12.93 |
| 80 | 12.16 | 15.84 | 19.03 | 11.84 | 14.94 | 19.19 |
| 90 | 15.65 | 23.03 | 22.72 | 14.80 | 21.17 | 29.10 |
| 100 | 19.30 | 36.50 | 27.13 | 17.94 | 32.20 | 47.71 |

***Table 1.*** *The spectral changes extracted from Fig. 5a-c. "Au NPs" refer to the bare "5 nm" sample, while "gr/Au NPs" refer to the corresponding graphene-covered sample.*

In the following we analyze in more details the vapour concentration-dependent optical response curves of the bare Au NPs. Careful examination reveals pronounced increases in the slopes of these curves, as they are emphasized in Fig. 6a.

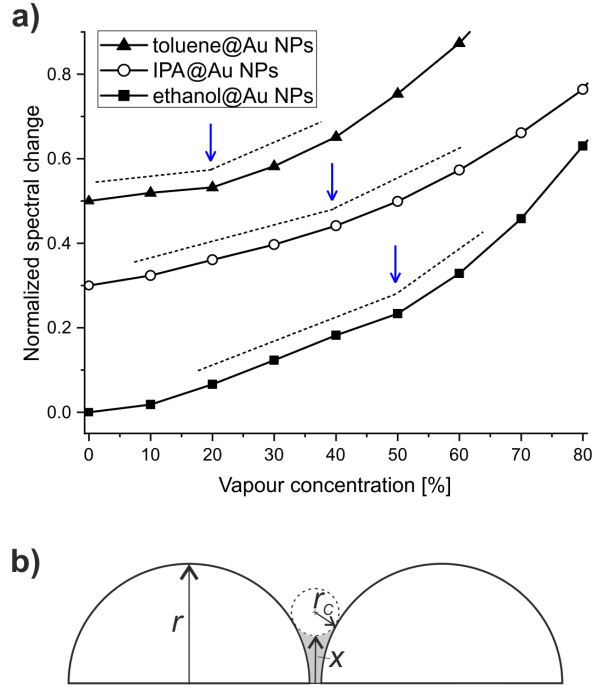

*Figure 6.* (a) Normalized spectral change of "5 nm" Au NPs sample during exposure to toluene (triangle), IPA (circle), and ethanol (square). The increases in the slopes are marked by arrows and dashed lines, as guides for the eye. For clarity, the curves were shifted vertically. (b) Model of capillary-condensate (grey) between two hemispherical nanoparticles.

Note that the increased slopes emerge at different concentration values for the three analytes: 20% for toluene, 40% for IPA, and 50% for ethanol. We discuss these values in the framework of capillary condensation, as underlying mechanism for the increase of the effective refractive index. A liquid trapped between two nanoparticles has a curved surface with a reduced vapour pressure ($P_0^C$), described by the Kelvin equation [36]:

$$RT \cdot \ln\frac{P_0^C}{P_0} = \gamma V_m \left(\frac{1}{x} - \frac{1}{r_C}\right), \qquad (2)$$

where $P_0$ is the standard vapour pressure, $R, T, \gamma, V_m$ are the universal gas constant, absolute temperature, surface tension, and molar volume of the condensed liquid, respectively. The two principal radii which characterize the liquid-vapour interface are denoted by $x$ and $r_C$ (see also Fig. 6b). We interpret the onset of the increased slopes as the start of condensation in the pores defined by the average interparticle spacing. For toluene, the onset is at $\left(\frac{P_0^C}{P_0}\right)_{toluene} = 0.2$ (20% vapour concentration), and from Eq. (2) we obtain $\left(\frac{1}{x} - \frac{1}{r_C}\right) = -1.34\ nm^{-1}$. Here we used $\gamma_{toluene} = 27.93 \times 10^{-3} N/m$, and $V_m^{toluene} = 0.10627\ m^3/kmol$ [37]. Due to the large spot-size of the light – compared to the size of NPs – used in the measurements, the obtained data is effectively averaged over the different particle spacings. Hence, we use the same $(1/x - 1/r_C)$ value to calculate the vapour pressure ratios at which condensation occurs for IPA and for ethanol. Using $\gamma_{IPA} = 20.93 \times 10^{-3} N/m$, $V_m^{IPA} = 0.07646\ \frac{m^3}{kmol}$, $\gamma_{ethanol} = 21.97 \times 10^{-3} N/m$, and $V_m^{ethanol} = 0.05839\ m^3/kmol$, we obtain $\left(\frac{P_0^C}{P_0}\right)_{IPA} = 0.42$ and $\left(\frac{P_0^C}{P_0}\right)_{ethanol} = 0.5$ for IPA and ethanol, respectively. These values are in very good agreement with the corresponding slope onsets (Fig. 6a), and show that the increased slopes can be attributed to capillary condensation in between the Au NPs. Note that the term $(1/x - 1/r_C)$ can yield the same numerical value for both $x \ll r_C$, and $x \gg r_C$. The first case corresponds to interparticle distances larger than $r$, the radius of NPs, while the latter case stands for closely spaced nanoparticles. In Fig. 1a we can observe examples of NP configurations for both cases. For IPA, we observe a second increase in the slope at 80% vapour concentration (Fig. 4c), which should be related to condensation in larger pores. When graphene is transferred onto the Au NPs, the pore structure of the sample changes. The graphene coverage of the sample is 40-50%, meaning that a large number of interparticle spacings are masked from the analytes. On the other hand, new, larger pores are formed by the suspended and curved graphene. As a result, for toluene we observe a slope change at increased vapour concentration (30%, Fig. 4d). Moreover, for IPA and ethanol we can identify clear slope changes only at high concentrations, 70% and 80%, respectively (Fig. 4b-c), also indicating larger pore dimensions for the graphene covered sample.

In order to better understand the optical response curves of the analytes, and in particular the lower spectral change observed for toluene exposure, one should also consider the interaction between the solvent molecules and graphene. IPA and ethanol are polar molecules, which bind to graphene non-covalently, with adsorption energies of -7.9 kcal/mol for ethanol [38], and around -10.1 kcal/mol for IPA [39]. Toluene, on the other hand, is a non-polar molecule, also binding to graphene through physisorption. The corresponding adhesion energy is -15.1 kcal/mol [38]. The larger adhesion energy shows that toluene is prone to adhere to graphene more than IPA or ethanol, which apparently contradicts the smaller spectral change in the presence of toluene vapour. Density functional theory (DFT) calculations by Pinto *et al.* [40] show that toluene adsorbes flat against the graphene, preferably in *AB* stacking configuration. Patil and Caffrey [39] showed recently by DFT that significant charge reorganization occurs on both the adsorbed molecule and graphene, although negligible charge transfer is involved. Toluene induces local electron density depletion, while IPA induces electron density accumulation in graphene, at the adsorption site [39]. This induces changes also in the electron density of the underlying Au NPs due to strong electromagnetic coupling with graphene [41]. Thus, the opposite charge redistribution in graphene occuring for toluene and IPA can result in slight, opposite shifts of the LSPR of the Au NPs as well [42]. This interaction effect adds up to the LSPR shift based on refractive index change. In particular, we think that at low toluene concentration (<20%) the refractive index based LSPR shift dominates, while interaction effects become important at somewhat higher concentrations (≥30%) where capillary condensation occurs. This could explain the change in the sensitivity of the graphene-covered Au NPs observed for toluene vapour. However, this calls for detailed theoretical calculations where the interactions between adsorbate, corrugated graphene, and the Au NPs are all considered, which goes beyond the scope of the current work.

## 4 Conclusions

Graphene-covered gold nanoparticles were produced and their vapour sensing properties were investigated by measuring the LSPR shift of the Au NPs. We found that smaller, dome-like Au NPs were more sensitive to ethanol, IPA, and toluene vapours compared to slightly larger, flat NPs. The slope changes observed on the optical response curves of dome-like Au NPs could be well described by capillary condensation. The fast response and recovery of Au NPs were preserved on the graphene-covered samples as well. We demonstrated that the presence of a corrugated graphene overlayer increased the sensitivity to ethanol and IPA, while it decreased it towards toluene exposure (at concentrations ≥30%). Nevertheless, at low toluene concentrations (10%) where capillary condensation does not yet occur, the graphene covered Au NPs are more sensitive to toluene, compared to bare Au NPs. The detection mechanism based on refractive index change does not fully explain the induced LSPR shifts. The interactions between adsorbate, corrugated graphene, and the Au NPs have to be considered, which requires further theoretical investigations. As a perspective, sensor arrays can be applied including graphene-covered and bare Au NPs as well, which can increase both the sensitivity and the selectivity of the hybrid system, offering the possibility of "fingerprinting" organic vapours.

**Conflicts of interest**

There are no conflicts of interest to declare.

**Acknowledgements**

The Authors acknowledge financial support from the National Research, Development and Innovation Office (NKFIH) in Hungary, through the Grants K_119532, K_111741, K_115724, FK_128327, KH_129578 and KH_129587, as well as from the Korea-Hungary Joint Laboratory for Nanosciences. The research leading to these results was supported by the ÚNKP-18-3 New National Excellence Program of the Ministry of Human Capacities and has received funding from the People Programme (Marie Curie Actions) of the European Union's Seventh Framework Programme under REA grant agreement no. 334377.

**Notes and references**

[1] A. K. Geim and K. S. Novoselov, *Nat. Mater.*, 2007, **6**, 183–191.

[2] M. J. Allen, V. C. Tung and R. B. Kaner, *Chem. Rev.*, 2010, **110**, 132–145.

[3] A. Ghosh, K. S. Subrahmanyam, K. S. Krishna, S. Datta, A. Govindaraj, S. K. Pati and C. N. R. Rao, *J. Phys. Chem. C*, 2008, **112**, 15704–15707.

[4] H. J. Yoon, D. H. Jun, J. H. Yang, Z. Zhou, S. S. Yang and M. M. C. Cheng, *Sens. Actuators, B*, 2011, **157**, 310–313.

[5] Q. He, S. Wu, Z. Yin and H. Zhang, *Chem. Sci.*, 2012, **3**, 1764–1772.

[6] U. Latif and F. L. Dickert, *Sensors*, 2015, **15**, 30504–30524.

[7] T. Wang, D. Huang, Z. Yang, S. Xu, G. He, X. Li, N. Hu, G. Yin, D. He and L. Zhang, *Nano-Micro Lett.*, 2016, **8**, 95–119.

8 K. Xu, C. Fu, Z. Gao, F. Wei, Y. Ying, C. Xu and G. Fu, *Instrum. Sci. Technol.*, 2018, **46**, 115–145.

9 A. Nag, A. Mitra and S. C. Mukhopadhyay, *Sensors Actuators, A*, 2018, **270**, 177–194.

10 C.-S. Cheng, Y.-Q. Chen and C.-J. Lu, *Talanta*, 2007, **73**, 358–365.

11 S. Underwood and P. Mulvaney, *Langmuir*, 1994, **10**, 3427–3430.

12 M. D. Malinsky, K. L. Kelly, G. C. Schatz and R. P. Van Duyne, *J. Am. Chem. Soc.*, 2001, **123**, 1471–1482.

13 K. M. Mayer and J. H. Hafner, *Chem. Rev.*, 2011, **111**, 3828–3857.

14 P. T. Yin, T.-H. Kim, J.-W. Choi and K.-B. Lee, *Phys. Chem. Chem. Phys.*, 2013, **15**, 12785–12799.

15 P. T. Yin, S. Shah, M. Chhowalla and K.-B. Lee, *Chem. Rev.*, 2015, **115**, 2483–2531.

16 S. Bai and X. Shen, *RSC Adv.*, 2012, **2**, 64–98.

17 S. Basu and S. Hazra, *C*, 2017, **3**, 29.

18 C. Tan, X. Huang and H. Zhang, *Mater. Today*, 2013, **16**, 29–36.

19 R. Alharbi, M. Irannejad and M. Yavuz, *Sensors*, 2019, **19**, 862.

20 S. Szunerits, Q. Wang, A. Vasilescu, M. Li and R. Boukherroub, in *Nanocarbons for Electroanalysis*, ed. S. Szunerits, R. Boukherroub, A. Downard and J.-J. Zhu, John Wiley & Sons, Ltd, Chichester, UK, 2017, Chapter 6, 139–172.

21 I. Khalil, N. Julkapli, W. Yehye, W. Basirun, S. Bhargava, I. Khalil, N. M. Julkapli, W. A. Yehye, W. J. Basirun and S. K. Bhargava, *Materials (Basel)*, 2016, **9**, 406.

22 B. Mehta, K. D. Benkstein, S. Semancik and M. E. Zaghloul, *Sci. Rep.*, 2016, **6**, 21287.

23 D. W. Boukhvalov, *Surf. Sci.*, 2010, **604**, 2190–2193.

24 A. Pálinkás, P. Süle, M. Szendrő, G. Molnár, C. Hwang, L. P. Biró and Z. Osváth, *Carbon*, 2016, **107**, 792–799.

25 H. Jans and Q. Huo, *Chem. Soc. Rev.*, 2012, **41**, 2849–2866.

26 Z. Osváth, A. Deák, K. Kertész, G. Molnár, G. Vértesy, D. Zámbó, C. Hwang and L. P. Biró, *Nanoscale*, 2015, **7**, 5503–5509.

27 R. Nicolas, G. Lévêque, P.-M. Adam and T. Maurer, *Plasmonics*, 2018, **13**, 1219–1225.

28 P. K. Jain, W. Huang and M. A. El-Sayed, *Nano Lett.*, 2007, **7**, 2080–2088.

29 G. Longobucco, G. Fasano, M. Zharnikov, L. Bergaminic, S. Corni, M. A. Rampi, *Sensors and Actuators B*, 2014, **191**, 356–363.

30 S. P. Scheeler, S. Mühlig, C. Rockstuhl, S. B. Hasan, S. Ullrich, F. Neubrech, S. Kudera and C. Pacholski, *J. Phys. Chem. C*, 2013, **117**, 18634–18641.

31 K. A. Willets and R. P. Van Duyne, *Annu. Rev. Phys. Chem.*, 2007, **58**, 267–297.

32 J. N. Anker, W. P. Hall, O. Lyandres, N. C. Shah, J. Zhao and R. P. Van Duyne, *Nat. Mater.*, 2008, **7**, 442–453.

33 Christian Reichardt, Solvents and Solvent Effects in Organic Chemistry, Wiley-VCH Publishers, 3rd ed., 2003

34 S. Cheon, K. D. Kihm, H. goo Kim, G. Lim, J. S. Park and J. S. Lee, Sci. Rep., 2014, **4**, 6364.

35 B. G. Ghamsari, J. Tosado, M. Yamamoto, M. S. Fuhrer and S. M. Anlage, *Sci. Rep.*, 2016, **6**, 34166.

36 H.-J. Butt, K. Graf, M. Kappl, in Physics and Chemistry of Interfaces, ed. U. Krieg, Wiley-VCH Verlag GmbH & Co. KGaA, Weinheim, 2003, 16–22.

37 David R. Lide, ed., *CRC Handbook of Chemistry and Physics, Internet Version 2005*, CRC Press, Boca Raton, FL, 2005.

38 P. Lazar, F. Karlický, P. Jurečka, M. Kocman, E. Otyepková, K. Šafářová and M. Otyepka, J. Am. Chem. Soc., 2013, **135**, 6372–6377.

39 U. Patil and N. M. Caffrey, *J. Chem. Phys.*, 2018, **149**, 094702.

40 H. Pinto, R. Jones, J. P. Goss and P. R. Briddon, *Phys. Rev. B*, 2010, **82**, 125407.

41 J. Niu, Y. J. Shin, J. Lee, J. H. Ahn and H. Yang, *Appl. Phys. Lett.*, 2012, **100**, 1116–1120.

42 M. Cittadini, M. Bersani, F. Perrozzi, L. Ottaviano, W. Wlodarski and Alessandro Martucci, Carbon, 2014, **69**, 452–459.